%% file: sample-manuscript.tex
\documentclass[acmsmall,screen]{acmart}
\usepackage{microtype}
\usepackage{graphicx}
\usepackage{subfigure}
\usepackage{booktabs} %
\usepackage{amsmath,amsfonts}
\usepackage{textcomp}
\usepackage{xcolor}
\usepackage{makecell}
\usepackage{multirow}
\usepackage{diagbox}
\usepackage{balance}
\usepackage[flushleft]{threeparttable}
\usepackage[utf8]{inputenc} 
\usepackage[T1]{fontenc}    
\usepackage{hyperref}       
\usepackage{url}            
\usepackage{booktabs}      
\usepackage{amsfonts}       
\usepackage{nicefrac}      
\usepackage{microtype}
\usepackage{multirow}
\usepackage{listings}
\usepackage{adjustbox}
\usepackage{diagbox}
\usepackage[ruled, vlined, linesnumbered]{algorithm2e}
\usepackage{booktabs}
\usepackage{enumitem}
\usepackage{framed}

\usepackage{pifont}

\newcommand{\technique}{DeCon}

\AtBeginDocument{%
  \providecommand\BibTeX{{%
    \normalfont B\kern-0.5em{\scshape i\kern-0.25em b}\kern-0.8em\TeX}}}

\begin{document}

\title{DeCon: Detecting  Incorrect Assertions via Postconditions Generated by a Large Language Model}

\author{Hao Yu}
\affiliation{%
  \institution{Peking University, Beijing}
  \country{China}}
\email{yh0315@pku.edu.cn}

\author{Tianyu Chen}
\affiliation{%
  \institution{Peking University, Beijing}
  \country{China}}
\email{}

\author{Jiaming Huang}
\affiliation{
  \institution{Peking University, Beijing}
  \country{China}}
\email{}

\author{Zongyang Li}
\affiliation{
  \institution{Peking University, Beijing}
  \country{China}}
\email{}

\author{Dezhi Ran}
\affiliation{
  \institution{Peking University, Beijing}
  \country{China}}
\email{}

\author{Xinyu Wang}
\affiliation{
  \institution{University of Michigan at Ann Arbor}
  \country{USA}}
\email{}

\author{Ying Li}
\affiliation{
  \institution{Peking University, Beijing}
  \country{China}}
\email{li.ying@pku.edu.cn}

\author{Assaf Marron}
\affiliation{
  \institution{Dept. of Computer Science and Applied Mathematics Weizmann Institute of Science}
  \country{Rehovot, Israel}}
\email{}

\author{David Harel}
\affiliation{
  \institution{Dept. of Computer Science and Applied Mathematics Weizmann Institute of Science}
  \country{Rehovot, Israel}}
\email{}

\author{Yuan Xie}
\affiliation{
  \institution{The Hong Kong University of Science and Technology}
  \country{China}}
\email{}

\author{Tao Xie}  
\affiliation{
  \institution{Key Lab of HCST (PKU), MOE; SCS; Peking University}
  \country{China}}
\email{taoxie@pku.edu.cn}

\renewcommand{\shortauthors}{Yu, et al.}

\begin{abstract}
Recently, given the docstring
for the target problem and the target function signature, large language models (LLMs) have been used not only to generate source code, but also to generate test cases, consisting of test inputs and assertions (e.g., in the form of checking an actual output against the expected output). However, as shown by our empirical study on assertions generated by four LLMs for the HumanEval benchmark, over 62\% of the generated assertions are incorrect (i.e., failed on the ground-truth problem solution).
To detect incorrect assertions (given the docstring
 and the target function signature along with a sample of example inputs and outputs), in this paper, we propose a new approach named \technique{} to effectively detect incorrect assertions via LLM-generated  postconditions for the target problem (a postcondition is a predicate that must always be true just after the execution of the ground-truth problem solution).
Our approach requires a small set of I/O examples (i.e., a sample of example inputs and outputs) for the target problem (e.g., the I/O examples included in the docstring for a target problem in HumanEval).
We use the given I/O examples to filter out those  LLM-generated postconditions that are violated by at least one given I/O example. We then  use the remaining  postconditions to detect incorrect assertions as those assertions that violate at least one remaining postcondition.
Experimental results show that \technique{} can detect averagely more than 64\% (63\% and 65.5\% detected by GPT-3.5 and GPT-4, respectively) incorrect assertions generated by four state-of-the-art LLMs, and \technique{} can also improve the effectiveness of these LLMs in code generation by 4\% in terms of Pass@1.
In addition, although \technique{} might filter out correct assertions, the fault-finding ability of the remaining correct assertions decreases only  slightly.

\end{abstract}



\maketitle
\input{intro}

\input{background}
\input{approach}

\input{evaluation}

\input{experiments}

\input{discussion}
\input{threats}
\input{relatedwork}
\input{conclusion}

\bibliographystyle{ACM-Reference-Format}
\bibliography{sample-base}

\appendix

\end{document}

%% file: intro.tex
\section{introduction}
\label{sec.intro}

Software testing can be used to validate the correctness of a program under test. 
To conduct software testing, developers write test cases, consisting of test inputs and assertions, to detect faults for preventing software failures at deployment time~\cite{kochhar2015code,vahabzadeh2015empirical}. To reduce the manual effort involved in writing test cases, 
various tools have been proposed to automate the generation of test cases, e.g., EvoSuite ~\cite{Evosuite}, JBSE~\cite{jbse2016Braione}, and Randoop~\cite{Randoop24}.
However, the test cases generated by these  tools help detect only crashing faults or regression faults, but are incapable of detecting non-crashing faults, e.g., logic-related faults.
For example, Randoop~\cite{Randoop24} and EvoSuite~\cite{Evosuite} create assertions based on capturing and asserting the return values of all non-void-return methods in the generated test inputs.  EvoSuite additionally  reduces these assertions based on mutation testing~\cite{mutation1,mutation2}.

Although recent approaches~\cite{deepAssert,dinalla2022toga,yu2022automated,Tufano2022assert} based on deep learning (DL) can generate assertions for detecting non-crashing and non-regression faults, these approaches face three major limitations.
First, these approaches have \textbf{limited generalization ability}. These approaches are trained on limited training sets, with limited training data and programming languages, and thus are often not effective when  applied on other datasets. 
Second, these approaches achieve \textbf{limited effectiveness}. A study~\cite{Tufano2022assert} has shown that the top-1 accuracy of these  approaches is only 26.40\%. 
Third, these approaches suffer from \textbf{high false-positive rate}. A recent study~\cite{hossain2023neural} shows that over 47\% of the assertions generated by TOGA~\cite{dinalla2022toga} (a state-of-the-art assertion generation approach) are false positives, i.e., incorrect assertions. Incorrect assertions can cause to report a correct program under test as faulty, greatly increasing debugging burden on developers.


To attempt to address the preceding limitations, recent efforts~\cite{Tufano2022assert,fakhoury2024exploring,xiong2023program,huang2023agentcoder,mali2024chiraag,yuan2023no,siddiq2023exploring,schafer2023empirical,el2024using} have leveraged large language models (LLM) to generate assertions,  achieving much higher effectiveness than DL-based approaches~\footnote{We refer DL-based approaches in our paper as non-LLMs DL-based approaches.} but still suffering from high false positives (i.e., a high percentage of generated assertions being incorrect assertions). 
%
For example, we conduct an empirical study (shown in Section~\ref{sec: assertion_gen}) on the quality of assertions generated by four popular LLMs (CodeGen~\cite{nijkamp2022conversational}, InCoder~\cite{fried2022incoder}, Codex~\cite{codex}, and GPT-3.5~\cite{chatgpt} ) on HumanEval~\cite{humanevaldata}, a widely used code generation benchmark. 
%
%
%
We measure the percentage  of correctly executed, incorrectly executed (i.e., failed on the ground-truth problem solution), and non-executable assertions (e.g., unable to be successfully parsed) among all the generated assertions, with the last two parts being incorrect assertions. 
The study shows that the percentage of incorrectly executed assertions among all the generated assertions is 54.1\%, and the percentage of non-executable assertions is 8.3\%. In other words, the percentage of incorrect assertions among all the generated assertions is 62.4\%.


To effectively detect incorrect assertions, in this paper, we propose a new approach named \technique{} that uses an LLM to generate postconditions to detect incorrect assertions. 
Besides the docstring and the target function signature for the target problem, our approach requires a small set of I/O examples (i.e., a sample of example inputs and outputs) for the target problem; these I/O examples can be manually prepared or can be derived by manually confirming a generated expected output for a test input.
Developers can easily write a few I/O examples additionally for the target problem (e.g., the I/O examples included in the docstring for a target problem in HumanEval).
Note that these I/O examples cannot be directly used to detect incorrect assertions, so we use the I/O examples in the docstring written by developers to reduce incorrect postconditions generated by an LLM and thus detect incorrect assertions with the remaining postconditions.

In particular, DeCon includes three steps. First, we feed the given function signature and docstring for the target problem to an LLM to generate candidate postconditions.
In this step, we design a prompt format for the LLM in order to generate formal and executable postconditions.
Second, among the candidate postconditions, we use the given
I/O examples to filter out those candidate postconditions
that are violated by at least one given I/O example. 
Third, we use the remaining postconditions to detect incorrect assertions as
those assertions that violate at least one remaining postcondition.

Experimental results show that \technique{} can detect more than 63.0\% (precision) and 65.5\% (precision) incorrect assertions generated by four LLMs (CodeGen~\cite{nijkamp2022conversational}, InCoder~\cite{fried2022incoder}, Codex~\cite{codex}, and GPT-3.5~\cite{chatgpt}) with the postconditions generated by GPT-3.5 and GPT-4, respectively. 
At the same time, only 9.7\% (recall being 90.3\%) and 6.5\% (recall being 93.5\%) of correct assertions are misjudged as incorrect assertions. 
After combining Recall and Precision, the F1 score is 73.8\% and 76.5\% for GPT-3.5 and GPT-4, respectively.
After removing the detected incorrect assertions, \technique{} can improve the average Pass@1 of the four LLMs in code generation by 3.2\% and 4.4\% 
with the postconditions generated by GPT-3.5 and GPT-4, respectively.
Although \technique{} might filter out correct assertions, the remaining correct assertions can still retain  99.3\% 
 fault-finding ability.
Compared to not using the I/O examples to filter out incorrect postconditions generated by an LLM, using the I/O examples can help gain 73.9\% (96.4\%-22.5\%) and 25\% (99.3\%-74.3\%) higher fault-finding ability when faults are detected via LLM-generated assertions that do not violate any postcondition generated by GPT-3.5 and GPT-4, respectively. 

In summary, this paper makes the following main contributions:

\begin{itemize}
 
\item We conduct an empirical study on the quality of assertions generated by four LLMs on HumanEval and find that  62.4\% of generated assertions are incorrect.

\item We propose a new approach named \technique{} for detecting incorrect assertions by generating postconditions.

\item \technique{} not only detects an average of more than 64\% (63\% and 65.5\% detected by GPT-3.5 and GPT-4, respectively) incorrect assertions generated by four LLMs but also further improves the effectiveness of an LLM in the tasks of code generation and fault finding.
The source code and experimental results of \technique{} are open-source~\cite{sourcedata}.
\end{itemize}

The remainder of this paper is organized as follows. Section~\ref{sec:background} introduces a motivating example. Section~\ref{sec:approach} details our \technique{} approach. 
Sections~\ref{sec:evalution} and~\ref{sec:experiments} describe the experimental setup and experimental results. 
Section~\ref{sec:discuss} discusses issues in our work.
Section~\ref{sec:threats} and ~\ref{sec:relatedwork} discuss the threats to validity and related work.
Section~\ref{sec:conclusion} concludes this paper.

%% file: background.tex
\section{Background and Motivating Example}\label{sec:background}
In this section, we first show the motivation example of our work.
Then, we detail several LLMs for code generation.
Finally, we discuss the importance of reducing incorrect assertions in the task of test generation.


\subsection{Motivating Example}
\begin{figure*}
	\centering
	\includegraphics[width=1.0\textwidth]{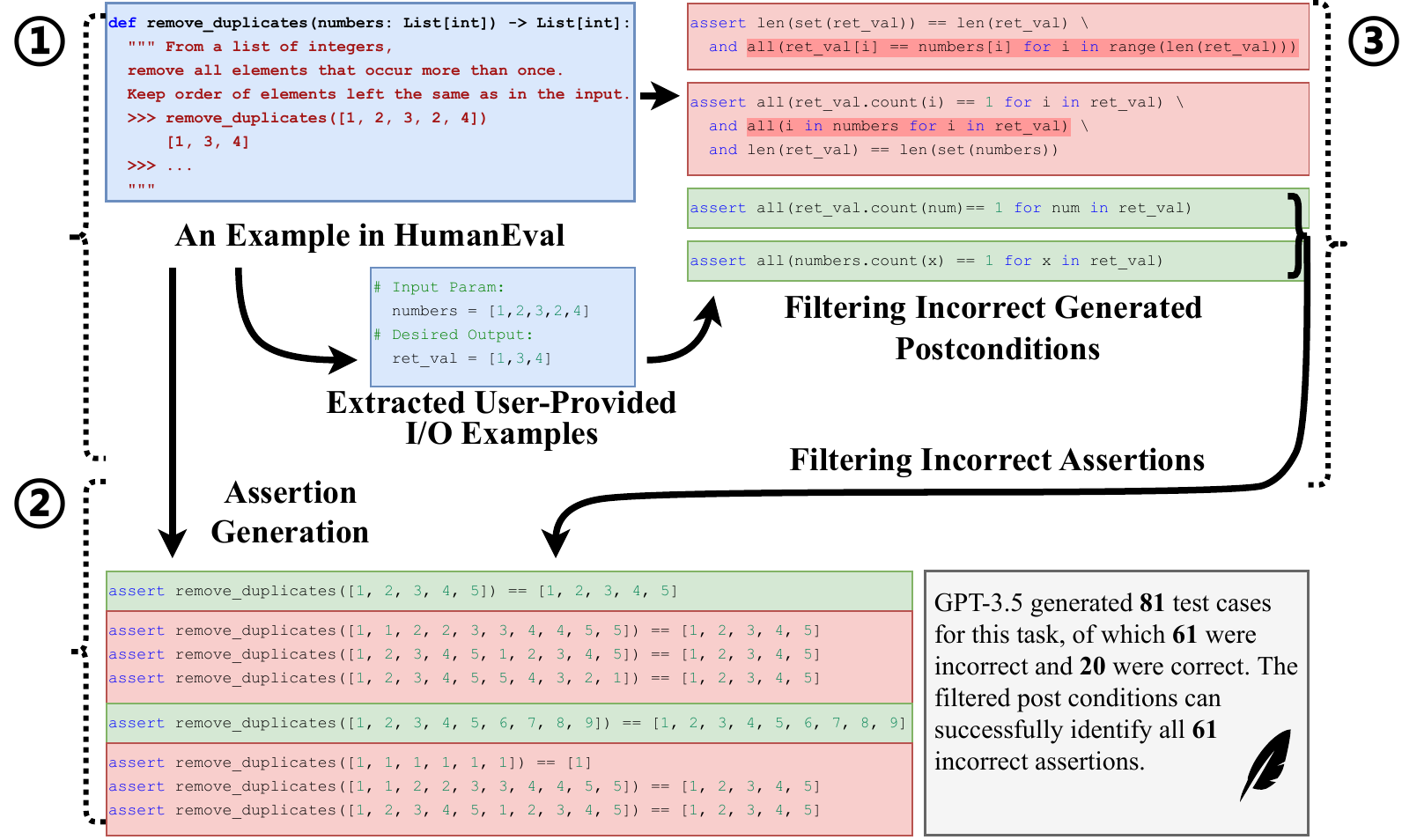}
	\caption{A motivating example of \technique{}}
	\label{fig: motivation}
\end{figure*}

Figure~\ref{fig: motivation} shows the motivating example of \technique{}.
Overall, the key idea of \technique{} is that \technique{} uses postconditions to filter out incorrect assertions.
As shown in Figure~\ref{fig: motivation}, when the developer provides a function signature and function description with one or two examples of input and output of the given function (part 1), the assertions generated by the tools of assertion generation for the given function signature and description are shown in part 2. The postconditions generated by LLMs for the given function signature and description are shown in part 3.
After generating assertions and postconditions, we automatically extract user-provided input and output examples from the function description.
Then, we use the user-provided input and output examples to filter the incorrectly generated postconditions.
The postconditions on the red background (filtered out) in part 2 violate the I/O examples, while the postconditions on the green background (retained) in part 2 do not violate the I/O examples.
Note that we can guarantee that the filtered-out postconditions are all incorrect, but we cannot guarantee that the remaining postconditions are all correct. The remaining postconditions may also misjudge correct assertions as incorrect ones.
After generating postconditions that do not violate the I/O examples, we use these postconditions to filter assertions that violate these postconditions. The assertions with a red background in part 3 are filtered out, while the assertions with a green background in part 3 are retained.

Finally, GPT-3.5 generated 81 assertions for this problem, of which 61 were incorrect and 20 were correct.
The remaining postconditions can successfully identify all 61 incorrect assertions and do not misjudge any correct assertions.

\subsection{LLMs for Code Generation}
CodeGen~\cite{nijkamp2022conversational} is a series of conversational text-to-code LLMs.
CodeGen used three-stage training to produce three models.
The first one produced CodeGen-NL, which was trained on a natural language dataset named The Pile~\cite{thepile}.
The second one produced CodeGen-Multi, which was further trained on a multiple-programming-language dataset named BigQuery. 
The third one produced CodeGen-Mono, which was built upon CodeGen-Multi with additional training on Python-only code.
InCoder~\cite{fried2022incoder} uses copyrighted source code for training and adds a mechanism to predict the current token to be generated using the following information.
Codex~\cite{codex} is the first work to use large generative pre-trained models to generate complete functions from natural language. 
After Microsoft presented Codex, DeepMind presented AlphaCode~\cite{alpha_code}, which specialized in programming contests and performed on par with median human developers.
CodeGPT~\cite{lu2021codexg} targets generating class member functions in Java, given a natural language description and class environment~\cite{iyer-etal-2018-mapping}. 
StarCoder~\cite{li2023starcoder} is a multilingual code generation model with a 15B parameter size. StarCoder crawled licensed code repositories from GitHub as pre-training data while removing sensitive personal identical information. The code generation capability of StarCoder in Python is on par with the model behind Copilot.
WizardCoder~\cite{luo2023wizardcoder} conducts instruction fine-tuning based on StarCoder, further improving the model's performance on the HumanEval benchmark.

Due to the training expected to include both source code and test code, these models can generate both source code and assertions.
In this work, we selected CodeGen, InCoder, Codex, and GPT-3.5 to generate assertions. 
The selected motivation is that CodeT's work has already generated assertions using CodeGen, InCoder, and Codex. We do not have the computational resources to generate assertions for other LLMs.

\subsection{The importance of detecting incorrect assertions in test generation}

According to our statistics, more than half of the assertions generated by the LLMs for HumanEval target problems are incorrect.
Detecting incorrect assertions generated by LLMs is crucial.
Our work is the first to point out how to reduce the false positives of the assertions automatically generated by LLMs.

%% file: approach.tex
\section{Approach}\label{sec:approach}

\begin{figure*}
	\centering
	\includegraphics[width=1.0\textwidth]{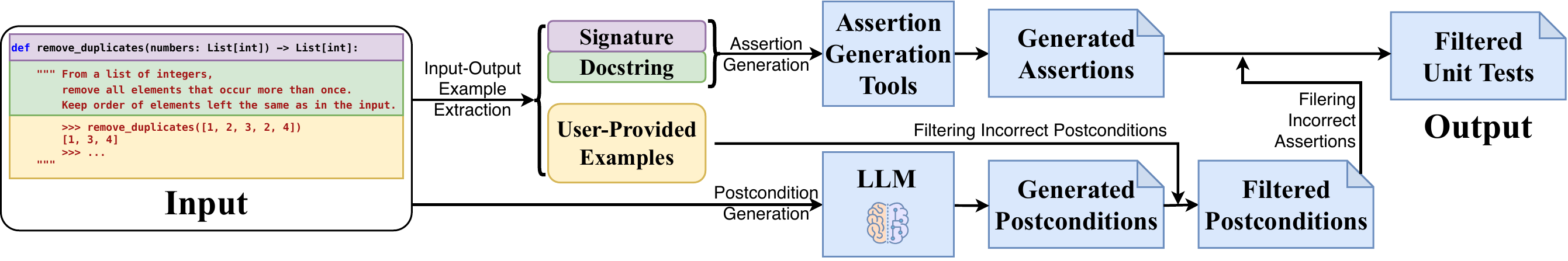}
	\caption{Overview of \technique{}}
	\label{fig: approachoverview}
\end{figure*}

\begin{algorithm}[t]
\footnotesize
	\SetKwData{eval}{\textbf{eval}}
    \SetKwData{not}{\textbf{not}}
	\SetKwFunction{compatible}{compatible}	
	\SetKwInOut{Input}{Input}\SetKwInOut{Output}{Output}
 
	\Input{ $signature$, an input method's signature.}
    \Input{ $doc$, the docstring of this method.}
    \Input{ $AssertionGen$, an LLM that generates assertions.}
    \Input{ $PostGen$, an LLM that generates postconditions.}
	\Output{$assertions$, a list of assertions for this method}
	\BlankLine

        \tcp{check whether one assertion and one postcondition are compatible}
        \SetKwProg{Fn}{Function}{ :}{end}
        \Fn{\compatible(assertion, condition)}{
        \tcp{an assertion is an (input, output) pair}
            $evalInput \gets assertion.input$\;
            $evalOutput \gets assertion.output$\;
            $postMethod \gets ``function(input, output) \{condition\}"$
            \BlankLine
            \Return{$\eval(postMethod(evalInput, evalOutput))$}
        }
        
        \BlankLine
        \tcp{Parsing docstring}
        $description \gets  doc.descriptions$\;
        $caseAssertions \gets doc.extractAssertions()$\;
        
        \BlankLine
        \tcp{generate assertions and postconditions}
        $candidateAsserts \gets  TestGen(signature, description)$\;
        $candidateConditions \gets PostGen(caseAssertions)$\;
        
        \BlankLine
        \tcp{filtering postconditions}
        $incConditions \gets \emptyset{}$\;
        \For{$condition \in candidateConditions$}{
            \For{$assertion \in caseAssertions$}{
                \If{$\not\ compatible(assertion, condition)$}{
                    $incConditions.add(condition)$\;
                }
            }
            $conditions \gets candidateConditions - incConditions$\;
        }
        
        \BlankLine
        \tcp{filtering assertions}
        $incAssertions \gets \emptyset{}$\;
        \For{$assertion \in candidateAsserts$}{
            \For{$condition \in conditions$}{
                \If{$\not\ compatible(assertion, condition)$}{
                    $incAssertions.add(assertion)$\;
                }
            }
            $assertions \gets candidateAsserts - incAssertions$\;
        }
        
        \Return{$assertions$}\;
\caption{Detecting Incorrect Assertions}\label{alg: overall}
\end{algorithm}

In this section, we first present an overview of \technique{} (Section~\ref{sec.overview}), then describe the detailed work in three parts: assertion generation (Section~\ref{sec: assertion_gen}), postcondition generation and filtering, (Section~\ref{sec: post_gen}) and incorrect assertion detection (Section~\ref{sec: assertion_reduction}).

\subsection{Overview}
\label{sec.overview}

Figure~\ref{fig: approachoverview} shows the overview of \technique{}.
The key idea of \technique{} is using postconditions to filter out incorrect assertions.
As shown in Figure~\ref{fig: approachoverview}, given a function signature and function description with one or two examples of input and expected output of the given function, the assertions generated by assertion generation tools for the given function signature and description.
We first extract the user-provided input and output from the docstring.
We extract an average of 2.87 I/O examples from the docstring of each problem in the HumanEval.
Then, we use LLMs to generate the postconditions for the given function signature and description.
Next, we use the user-provided input and output examples to filter the incorrect postconditions generated by LLMs.
Note that we can guarantee that the filtering out postconditions are all incorrect, but we cannot guarantee that the remaining postconditions are all correct. The remaining postconditions may also misjudge correct assertions as incorrect ones.
Finally, after generating postconditions that do not violate the I/O examples, we use these postconditions to filter assertions that violate these postconditions.



\input{approach/unitetestinggen}
\input{approach/postgen}
\input{approach/reducefalseassertion}

%% file: approach/unitetestinggen.tex
\subsection{Assertion Generation}\label{sec: assertion_gen}
The assertions that are generated by traditional tools (e.g., Evosuite and Randoop) are based on executing the program implementation; the assertions generated by traditional tools can never find logical bugs.

In this paper, we focus on only the assertions generated by LLMs. 
We do not use deep learning based assertion generation approaches to generate assertions for two reasons. First, current assertion generation approaches based on deep learning are based on a limited training set (training data and program language are all limited) and cannot be generalized to the HumanEval dataset. Second, a study~\cite{Tufano2022assert} has shown that the effectiveness of LLM-based assertion generation is higher than that of deep learning-based approaches.

Similar to existing work~\cite{chen2022codet}, we use the following prompt for LLMs to generate unit test cases:
``$description$ $of$ $the$ $function$ and check the correctness of $function\_name$ assert''.
We use a total of four LLMs (i.e., CodeGen, InCoder, Codex, and GPT3.5) to generate assertions (Lines 5 and 7 in Alg~\ref{alg: overall}).

%% file: approach/postgen.tex
\subsection{Postcondition Generation and Filtering}\label{sec: post_gen}

\subsubsection{User-Provided Examples Extraction}
\begin{figure}
	\centering
	\includegraphics[width=1.0\linewidth]{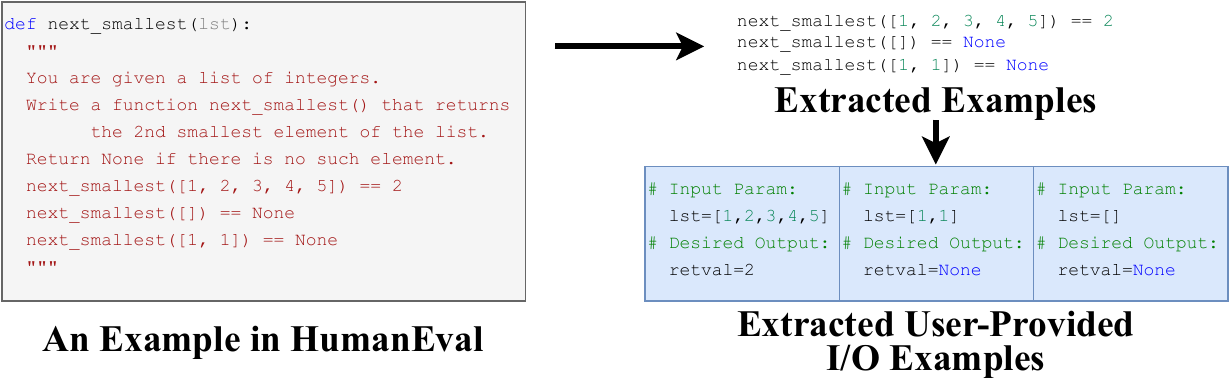}
	\caption{The processing of user-provided I/O examples extraction}
 \label{fig.ex1}
\end{figure}
Figure~\ref{fig.ex1} shows the processing of user-provided I/O examples extraction.
To automatically verify whether the postconditions generated by the LLM are correct for the I/O examples, we extract the I/O examples into the format of variable declarations (Line 6 in Alg~\ref{alg: overall}).

\subsubsection{Postcondition generation via LLMs}

Similar to existing work~\cite{endres2023formalizing}, we use the following prompt for GPT-3.5 and GPT-4 to generate postconditions:

``\textit{You have the following code context, function stub and natural language specification (in the form of a code comment) for \textbf{[FUNCTION NAME]}. When implemented, the function should comply with this natural language specification: \textbf{[ FUNCTION STUB, AND DOCSTRING HERE]}
Write a symbolic postcondition for \textbf{[FUNCTION NAME]} consisting of exactly one assert statement.
For variables, use only the function input parameters and a hypothetical return value, which we’ll assume is stored in a variable return\_val. If the post condition calls any functions external to the program context, they should only be those from the functional subset of \textbf{[PROGRAMMING LANGUAGE]}. 
By this, we mean functions that are pure (i.e., no side effects) such as \textbf{[PROGRAMMING LANGUAGE-SPECIFIC EXAMPLE]}.
Although the post condition should be less complex than the function itself, it should not be trivial. It should encapsulate an aspect of the function without implementing the function. The format of your response should be: code for exactly one postcondition with assert here.}''

In the preceding prompt, the bold parts represent inputs specific to different problems and programming languages, while the rest are fixed templates.
Since our experiment only focuses on the Python version of HumanEval, ``[PROGRAMMING LANGUAGE]'' is ``Python'', and ``[PROGRAMMING LANGUAGE-SPECIFIC EXAMPLE]'' is
``def fun (num1, num2): \textbackslash n \textbackslash t x=(num1 * num2)/num2 \textbackslash n \textbackslash t return x.''
For the example in the overview, ``[Function Name]'' is ``remove\_duplicates'', and ``[PROGRAM CONTEXT, FUSION STUB, AND DOCSTRING HER]'' is the first part of the overview.

\subsubsection{Filter postconditions with user-provided examples}
We extract each user-provided input and output from docstring as the variable declaration in a program, and put the generated postconditions after the variable declaration. 
We ensure that after combining the extracted input and output examples with the postconditions to be filtered into one Python file, the combined Python file is executable.
For any postcondition, if it fails to pass a user-provided example, it is considered that the postcondition generated is incorrect (Lines 10 to 15 in Alg~\ref{alg: overall}).

\begin{figure}
	\centering
	\includegraphics[width=1.0\linewidth]{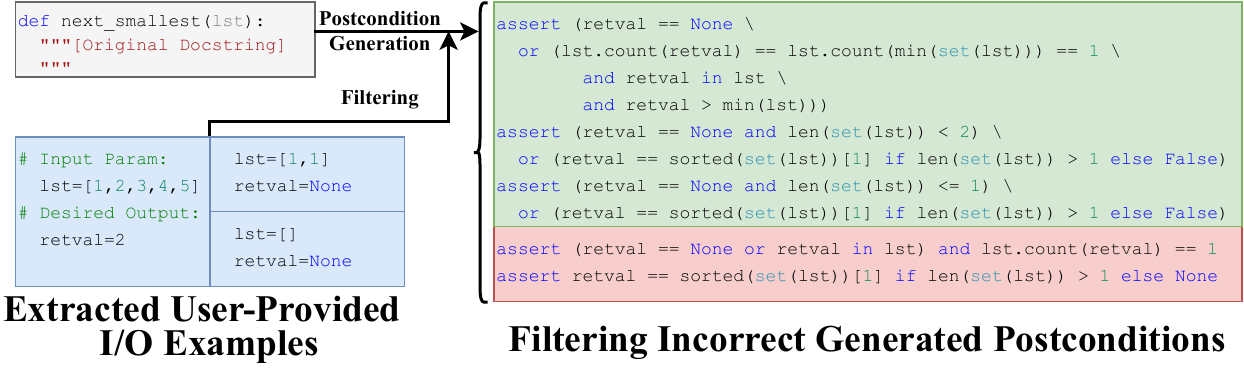}
	\caption{An example of filtering out incorrect postconditions}
 \label{fig.ex2}
\end{figure}

Figure~\ref{fig.ex2} shows an example of filtering out incorrect postconditions.
The three postconditions on the green background satisfy all three I/O examples, and the two postconditions on the red background do not satisfy at least one of the three I/O examples.

%% file: approach/reducefalseassertion.tex
\subsection{Incorrect Assertion Detection}
\label{sec: assertion_reduction}
\begin{figure}
	\centering
	\includegraphics[width=1.0\linewidth]{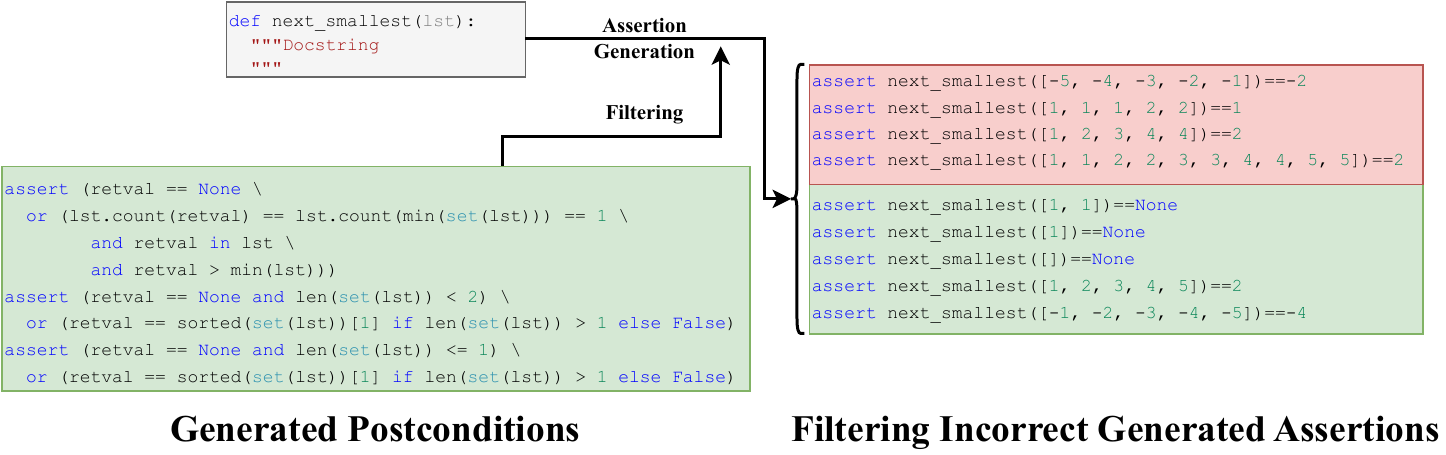}
	\caption{An example filtering out incorrect assertions}
 \label{fig.ex3}
 \vspace{-0.2cm}
\end{figure}
We extract each assertion as the variable declaration in a program, and put all filtered postconditions after the variable declaration. 
We ensure that after combining the extracted variables from the generated assertions with the filtered postconditions into one Python file, the combined Python file is executable.
For each assertion, if it fails to pass a postcondition, it is considered that the generated assertion is incorrect (Lines 16 to 21 in Alg~\ref{alg: overall}). 

Figure~\ref{fig.ex3} shows an example of filtering out incorrect assertions.
The five assertions on the green background satisfy all three remaining postconditions, and the four assertions on the red background do not satisfy at least one of the three filtered-out postconditions.

%% file: evaluation.tex
\section{evaluation}\label{sec:evalution}

In this section, we describe the setup of our experiment with four models (CodeGen, Incoder, Codex, and GPT-3.5) for detecting incorrect assertions in terms of research questions, evaluation dataset, model settings, and the evaluation metric.
Our experimental results are open-source~\cite{sourcedata}.

\subsection{Research Questions}

Our experiment answers the following research questions:
\begin{itemize}
    \item \textbf{RQ1:} What are the correctness of assertions generated by LLMs on HumanEval?

    \item \textbf{RQ2:} What are the correctness of postconditions generated by LLMs on HumanEval?

    \item \textbf{RQ3:} How effective is \technique{} in detecting incorrect assertions based on generated postconditions?

    \item \textbf{RQ4:} How does the incorrect postcondition filtering technique contribute to the effectiveness of detecting incorrect assertions?

    \item \textbf{RQ5:} How effective is \technique{} in improving the effectiveness of code generation for LLMs?

    \item \textbf{RQ6:} How effective is \technique{} in fault finding?

\end{itemize}


\subsection{Evaluation Dataset}

\subsubsection{\technique-AssertData}
We set the samples for each LLM to 100; each sample may contain one or more test cases, and each test case may contain one or more assertions.
We extracted all the assertions from the test cases generated by the LLMs and generated a total of 16,326, 126,529, 85,507, and 89,728 assertions for CodeGen, InCoder, Codex, and GPT-3.5, respectively.
After deduplicating these assertions, CodeGen, InCoder, Codex, and GPT-3.5 generated 13,970, 90,065, 54,643, and 29,452 assertions, respectively.



\subsubsection{HumanEval}
Released alongside Codex, HumanEval is a benchmark to measure code generation approaches on the functional correctness of programs synthesized from docstrings~\cite{codex}. It consists of 164 handwritten programming problems and solutions in Python, each of which includes a function signature, docstring, body, and several unit tests (7.7 tests per problem on average).
For each problem, the input contains two parts: one is an NL description of the problem, and another is the function signature of the solution that contains the function argument types and the expected return type.

\subsubsection{HumanEvalFix}
Muennighoff et al.~\cite{muennighoff2023octopack} manually injected faults into the standard implementation of each problem in HumanEval, resulting in incorrect implementations to obtain HumanEvalFix. All problems in HumanEvalFix are injected with fault. HumanEvalFix was initially intended to evaluate the effectiveness of program fixes, but due to its inclusion of correct and fault-prone code implementations, we use it to evaluate the fault-finding ability of assertions.

\subsection{Model settings}

\subsubsection{Assertion generation}
For all models (i.e., CodeGen, InCoder, Codex, and GPT-3.5), we generate 100 samples~\footnote{Note that each sample may contain one or more assertions} for each problem in HumanEval. The temperature and top-k are 0.8 and 0.95, respectively.
The kernel model we used for Codex, GPT-3.5, and GPT-4 is \textit{code-davinci-002}, \textit{GPT-3.5-Turbo}, and \textit{GPT-4-Turbo}.

\subsubsection{Postcondition generation}
For postcondition generation models (i.e., GPT-3.5 and GPT4), we generate five samples~\footnote{Note that each sample may contain one or more assert statements} for each problem in HumanEval. The temperature and top-k are the same as those of assertion generation models.

To avoid randomness in the experiment, we conducted the experiment three times when using GPT-3.5 and GPT4 to generate postconditions. In the end, we found that the results of the three experiments were relatively similar. Based on the weighted F1 scores, we selected the postconditions with the performance effect in the middle.

\subsection{Evaluation Metric}

\subsubsection{Metrics for incorrect assertions}

We use the correct implementation built-in in HumanEval to mark whether the assertions are correct.

\subsubsection{Metrics for compilability}
We use the ``return\_value'' of ``process'' in Python to determine whether there are syntax errors at runtime. If there are syntax errors, it is considered uncompilable.

\subsubsection{Metrics for detecting incorrect assertions}
Since we cannot guarantee that the generated postconditions and remaining postconditions (after filtering out incorrect postconditions by I/O examples) are completely correct, postconditions might classify the correct assertions as incorrect and also classify the incorrect assertions as correct. 
We choose Precision, Recall, and F1 scores to evaluate the effectiveness of \technique{} in detecting incorrect assertions.
These metrics are also adopted by existing work~\cite{dinalla2022toga,deiner2023automated} that generates assertions.
These metrics are defined as follows:
\begin{equation}~\label{eq: precision recall}
    \begin{aligned}
        &Precision = \frac{TP}{TP + FP},\quad Recall = \frac{TP}{TP + FN} \\
        & F1 = \frac{2 \times Precision \times Recall }{Precision + Recall}
    \end{aligned}
\end{equation}
where $TP$ and $FP$ are correct and incorrect assertions that are labeled as correct ones, and $TN$ and $FN$ are incorrect and correct assertions that are labeled as incorrect ones, respectively.



\subsubsection{Metrics for improving the effectiveness of code generation}
We adopt the Pass@K metric to evaluate the behavior correctness of generated code snippets according to assertions.
As we set \emph{n} (the number of samples) to 100 and calculate Pass@K for K in 1, 2, and 10, to avoid the issue of high sampling variance, we use the unbiased estimator of Pass@K implemented by Codex in HumanEval~\cite{humanevaldata}.

\subsubsection{Metrics for finding faults}
The implementation of each problem in the HumanEvalFix dataset has faults. The more problems with faults detected in the generated assertions, the higher the quality of the generated assertions.


%% file: experiments.tex
\section{Experiment Results and Analysis}\label{sec:experiments}
In this section, we show our experimental results and detail the analysis for each research question.

\input{evaluation/rq1}

\input{evaluation/rq2}
\input{evaluation/rq3}

\input{evaluation/rq4}

\input{evaluation/rq5}
\input{evaluation/rq6}

%% file: evaluation/rq1.tex
\subsection{\textbf{RQ1:} What is the correctness of assertions generated by LLMs on HumanEval?}\label{sec: assertion correctness}

We divide assertions generated by LLMs into the following three steps.
First, we remove duplicated assertions from all generated ones.
Second, we divide assertions generated by LLMs into two parts, compilable and non-compilable, according to whether one generated assertion can be successfully compiled.
Third, we divide compilable assertions into correct ones and incorrect ones (based on whether they are satisfied by the ground truth provided by the HumanEval dataset).

\begin{figure}
	\centering
	\includegraphics[width=0.8\linewidth]{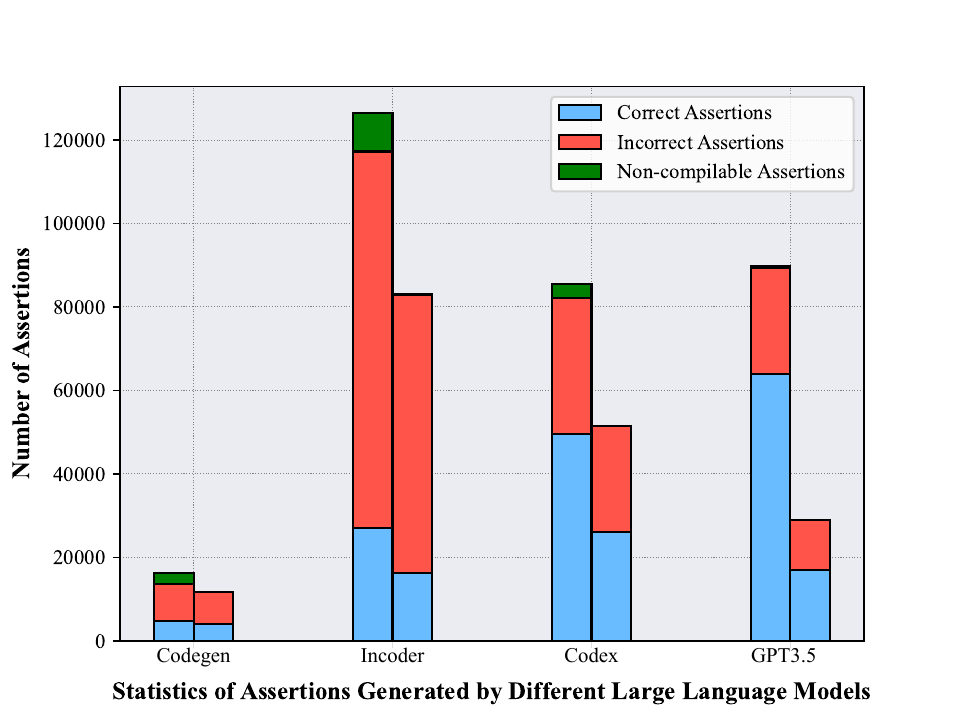}
	\caption{The number and proportion of different assertion types generated by different models }
 \label{fig.proportion_unique}
 \vspace{-0.2cm}
\end{figure}

Figure~\ref{fig.proportion_unique} shows the number and proportion of three assertion types generated by different LLMs.
For each LLM, the left column represents the results of assertions before deduplication, while the right column represents the results of assertions after deduplication.
For each column, the green sub-column represents the number of non-compilable assertions, the red sub-column represents the number of incorrect assertions, and the blue sub-column represents the number of correct assertions.


From Figure~\ref{fig.proportion_unique}, we find that: (1) \textbf{Duplication.} Repetition refers to the assertions being completely consistent at the string level. 
The proportions of duplicated assertions in CodeGen, InCoder, Codex, and GPT-3.5 are 14.4\%, 28.8\%, 36.1\%, and 67.2\%, respectively. 
(2) \textbf{Compilability.} 
We use the ```return\_value'' of ```process'' in Python to determine whether there are syntax errors at runtime. If there are syntax errors, it is considered uncompilable. We will replace compatibility with syntax correctness.
Before deduplication, non-compilable assertions generated by three LLMs (except CodeGen) count for 8.3\%, while the proportion of non-compilable assertions by all LLMs decreases substantially to less than 0.5\%, indicating that most non-compilable assertions are duplicated and easy to remove.
(3) \textbf{Correctness.}
After deduplication, the overall proportion of incorrect assertions among executable assertions on four LLMs is 59.6\% (54.1\% for all the generated assertions).
Specifically, the correct proportions of CodeGen, Incoder, Codex, and GPT-3.5 are 55.3\%, 74.2\%, 46.7\%, and 41.1\%, respectively.

Based on the component analysis of the assertions generated by four LLMs, we find that (1) deduplication and compilation can help generate unique and correct assertions. (2) There are still about 62.4\% incorrect assertions (8.3\% non-compilable assertions and 54.1\% compilable assertions) generated by LLMs, thus indicating the significance of detecting incorrect assertions.

\begin{figure}
	\centering
	\includegraphics[width=1.0\linewidth]{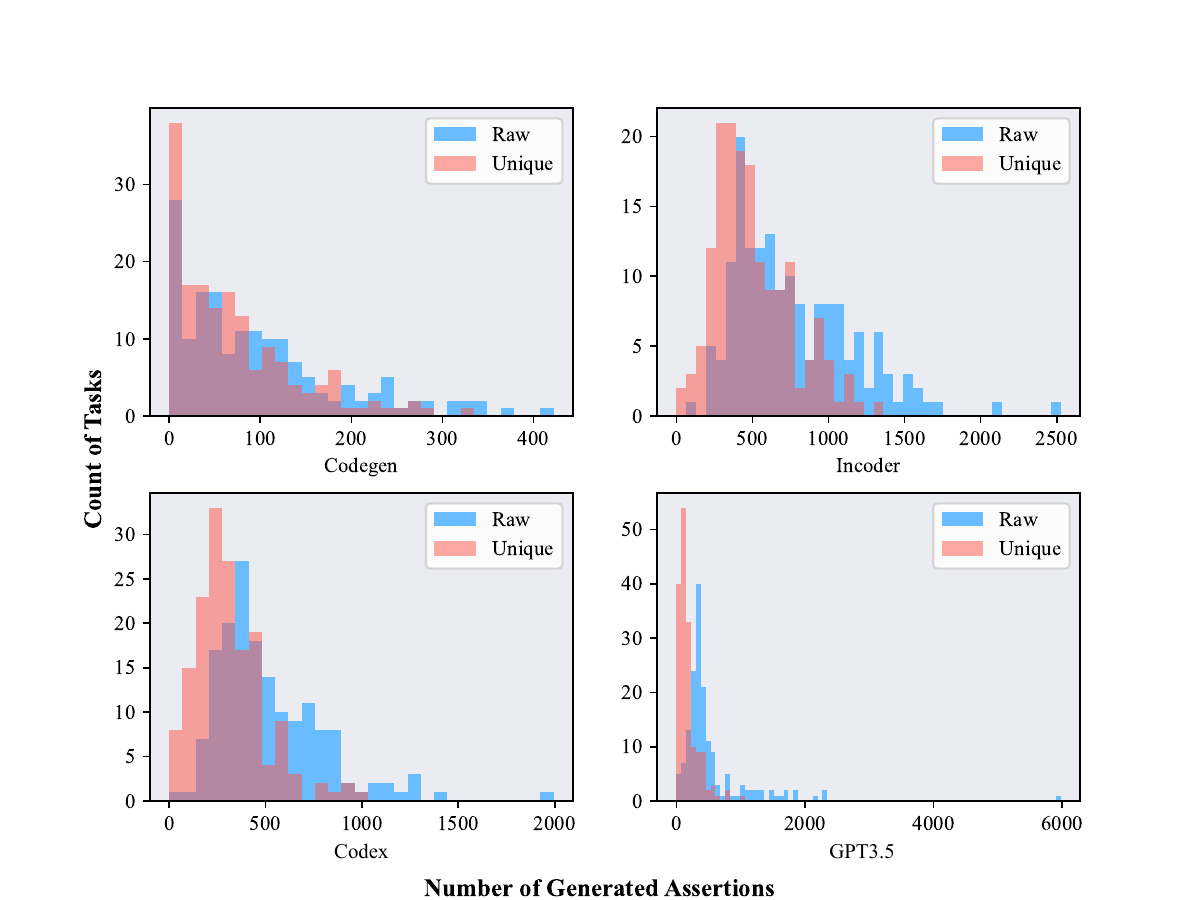}
	\caption{The number distributions of generated assertions over HumanEval's problems.}
    \label{fig.histplot_unique}
\vspace{-0.2cm}
\end{figure}

Additionally, Figure~\ref{fig.histplot_unique} shows the number distributions of generated assertions over HumanEval's problems.
Note that the number distributions of the four models vary substantially.
For example, the distribution of CodeGen looks like an exponential distribution (decreases only), while the distributions of the other three models are close to normal distributions (increase first and then decrease)~\cite{ahrens1972computer}.
Considering the substantial difference among various problems in RQ3 and RQ4, in addition to directly conducting evaluations on all assertions, we also designed another evaluation setting, conducting evaluations on the weighted results of each problem in HumanEval.

%% file: evaluation/rq2.tex
\subsection{\textbf{RQ2:} What is the correctness of postconditions generated by LLMs on HumanEval?}\label{sec:post correctness}


Since each postcondition may contain one or more assert statements,
GPT-3.5 generates a total of 1,900 assert statements for postconditions for 164 problems.
We short postconditions in the form of assert statements as postconditions in the rest of this paper.
After deduplication, GPT-3.5 generates a total of 1,698 postconditions. When using the built-in test cases of the HumanEval dataset for filtering, 671 incorrect postconditions are filtered out.
As for GPT-4, it generates a total of 820 postconditions for 164 problems.
After deduplication, GPT-4 generates a total of 671 postconditions.
When using the built-in test cases of the HumanEval dataset for filtering, 275 incorrect postconditions are filtered out.

Generally, we use the I/O examples in the docstring to filter out 539 incorrect postconditions among 671 incorrect postconditions (detected by the built-in test cases in the HumanEval) generated by GPT-3.5 and 194 incorrect postconditions among 275 incorrect postconditions (detected by the built-in test cases in the HumanEval) generated by GPT-4.
At the problem level, we find five problems where GPT-3.5 does not generate any correct postconditions and 16 problems whose generated postconditions are completely correct.
For GPT-4, there are 19 problems generated by GPT-4 that do not generate any correct postconditions and 87 problems whose generated postconditions are completely correct.

To explain the reason for GPT's less effectiveness in generating postconditions, we manually investigate the generated postconditions in 15 problems where GPT-3.5 can generate correct postconditions while GPT-4 fails.
We find that among these 15 problems, the postconditions generated by GPT-3.5 are quite simple ones, such as ``assert isinstance (return\_val, bool)'', and ``assert return\_val in ["YES", "NO"]''.

%% file: evaluation/rq3.tex

 \subsection{\textbf{RQ3:} How effective is \technique{} in detecting incorrect assertions based on generated postconditions?}\label{sec:detect1}

\begin{table*}[t]
\caption{Overall effectiveness of \technique{}}
\label{tab: overallwithfiltering}
\footnotesize
\begin{tabular}{cc|rrrrr|ccc|ccc}
\toprule
\multirow{2}{*}{\makecell[c]{AssertGen\\ Model}} & \multicolumn{1}{c}{\multirow{2}{*}{\makecell[c]{PostGen\\ Model}}} & \multicolumn{4}{c}{Number of assertions} & \multicolumn{1}{c|}{Raw}                             & \multicolumn{3}{c}{Unweighted Metrics} & \multicolumn{3}{c}{Weighted Metrics} \\
\cmidrule(lr){3-6}\cmidrule(lr){7-7}\cmidrule(lr){8-10}\cmidrule(lr){11-13}
& \multicolumn{1}{c}{}                               & \multicolumn{1}{|r}{TP}    & FP    & TN    & FN   & \multicolumn{1}{c|}{Prec.} & Prec.       & Rec.        & F1         & Prec.       & Rec.        & F1         \\ 
\cmidrule(lr){1-13}
\multirow{2}{*}{CodeGen}       & GPT-3.5                                            & 3507  & 2465  & 5262  & 417  & 0.337 & 0.587       & 0.894       & 0.709      & 0.591 & 0.734 & 0.655      \\
& GPT-4                                               & 3656  & 2257  & 5470  & 268  & 0.337 & 0.618       & 0.932       & 0.743      & 0.647 & 0.779 & 0.707      \\ 
\cmidrule(lr){1-13}
\multirow{2}{*}{InCoder}       & GPT-3.5                                            & 14317 & 15901 & 50932 & 1829 & 0.195 & 0.474       & 0.887       & 0.618      & 0.575 & 0.824 & 0.677      \\
& GPT-4                                               & 14601 & 15729 & 51104 & 1545 & 0.195 & 0.481       & 0.904       & 0.628      & 0.665 & 0.874 & 0.755      \\ 
\cmidrule(lr){1-13}
\multirow{2}{*}{Codex}         & GPT-3.5                                            & 23493 & 10391 & 15111 & 2499 & 0.505 & 0.693       & 0.904       & 0.785      & 0.732 & 0.871 & 0.795      \\
& GPT-4          & 24544 & 9823  & 15679 & 1448 & 0.505 & 0.714       & 0.944       & 0.813      & 0.788 & 0.917 & 0.848      \\ 
\cmidrule(lr){1-13}
\multirow{2}{*}{GPT-3.5}       & GPT-3.5                                            & 15715 & 4684  & 7407  & 1250 & 0.584 & 0.770       & 0.926       & 0.841      & 0.780 & 0.902 & 0.836      \\
& GPT-4                                               & 16325                      & 3950  & 8141  & 640  & 0.584                      & 0.805       & 0.962       & 0.877      & 0.815 & 0.941 & 0.873      \\
\bottomrule
\end{tabular}
\begin{itemize}
    \item The column ``Raw Prec.'' is defined as (TP + FN) / Total, the precision of assertion generated by LLMs
\end{itemize}
\vspace{-0.3cm}
\end{table*}

\subsubsection{Methodology}
Since we cannot guarantee that the generated postconditions and remaining postconditions (after filtering out incorrect postconditions by I/O examples) are completely correct, postconditions might classify the correct assertions as incorrect and also classify the incorrect assertions as correct. We evaluate the effectiveness of \technique{} in detecting incorrect assertions based on our generated postconditions from the perspective of three metrics: precision, recall, and F1 scores.
\textbf{As shown in Section~\ref{sec: assertion correctness}, we also average these three metrics among various problems as weighted metrics to mitigate various numbers of assertions in different problems.}
As shown in Figure~\ref{fig: approachoverview}, we conduct evaluations on each combination of AssertGen models (generating assertions) and PostGen models (generating postconditions).
Specifically, we employ CodeGen, InCoder, Codex, and GPT-3.5 as our AssertGen models and GPT-3.5 and GPT-4 as our PostGen models.



\subsubsection{Improvement in precisions}
Table~\ref{tab: overallwithfiltering} shows \technique{}'s precision, recall, and F1 scores in identifying correct assertions.
The results reveal that \technique{} can effectively improve the precision of assertions for each combination of AssertGen and PostGen models.
Notably, compare with the raw precision (the column ``Raw''), \technique{} boosts the precision of assertions by a minimum of 31.8\% (AssertGen model is GPT-3.5, PostGen model is GPT-3.5) and a maximum of 146.7\% (AssertGen model is InCoder, PostGen model is GPT-4), indicating that \technique{} can effectively improve the precision of assertions generated by LLMs by detecting incorrect assertions.

\subsubsection{Trade-offs between precision and recall}
Besides the improvement in precision, Table~\ref{tab: overallwithfiltering} also lists the results of recalls and F1 scores of \technique{}.
Notably, all the recall values exceed 0.887, indicating the \technique{}'s effectiveness in retaining correct assertions.
To reflect the trade-offs between precision and recall, we also use the F1 scores for evaluation, and \technique{} achieves an average F1 score of 0.752. 
Additionally, models with higher raw precision scores (e.g., Codex and GPT-3.5) can also benefit from \technique{}, thus indicating \technique{}'s generalization ability.

\subsubsection{Results of the weighted metrics}
Table~\ref{tab: overallwithfiltering} also shows the results weighted among various problems in HumanEval.
There is a slight increase in the precision (less than 0.1) and a minor decrease (less than 0.15) in recall when compared with unweighted metrics.
This difference demonstrates that \technique{} maintains consistent performance across different problems in the HumanEval dataset.

%% file: evaluation/rq4.tex
\subsection{RQ4: How does the incorrect postcondition filtering technique contribute to the effectiveness of detecting incorrect assertions?}\label{sec:detect2}

In the preceding section, we have shown the effectiveness of \technique{} in detecting incorrect assertions.
In this section, we conduct ablation studies to evaluate the contribution of incorrect postcondition filtering (i.e., filtering incorrect postconditions based on the I/O examples) to the achieved effectiveness.

\subsubsection{Methodology}
To evaluate the contribution of incorrect postconditions filtered by I/O examples, we conduct ablation studies that do not filter candidate postconditions based on the I/O examples extracted from docstrings (Lines 9-14 in Algorithm~\ref{alg: overall}).
We use precision, recall, and F1 scores in both weighted and unweighted settings. 

\begin{table*}[t]
\caption{Overall effectiveness of \technique{} without I/O example filtering for incorrect postconditions}
\label{tab: overallwithoutfiltering}
\footnotesize
\begin{tabular}{cc|rrrrr|ccc|ccc}
\toprule
\multirow{2}{*}{\makecell[c]{AssertGen\\ Model}} & \multicolumn{1}{c}{\multirow{2}{*}{\makecell[c]{PostGen\\ Model}}} & \multicolumn{4}{c}{Number of assertions} & \multicolumn{1}{c|}{Raw}                             & \multicolumn{3}{c}{Unweighted Metrics} & \multicolumn{3}{c}{Weighted Metrics} \\
\cmidrule(lr){3-6}\cmidrule(lr){7-7}\cmidrule(lr){8-10}\cmidrule(lr){11-13}
& \multicolumn{1}{c}{}                               & \multicolumn{1}{|r}{TP}    & FP    & TN    & FN   & \multicolumn{1}{c|}{Prec.} & Prec.       & Rec.        & F1         & Prec.       & Rec.        & F1         \\ 
\cmidrule(lr){1-13}
\multirow{2}{*}{CodeGen}       & GPT-3.5                        & 749    & 253  & 7474  & 3175  & 0.337 & 0.748       & 0.191       & 0.304      & 0.220        & 0.138        & 0.169      \\
 & GPT-4                           & 2523   & 784  & 6943  & 1401  & 0.337 & 0.763       & 0.643       & 0.698      & 0.601 & 0.55 & 0.574      \\
\cmidrule(lr){1-13}
\multirow{2}{*}{InCoder}       & GPT-3.5                        & 3131   & 904  & 65929 & 13015 & 0.195 & 0.776       & 0.194       & 0.310      & 0.305 & 0.166 & 0.215 \\
& GPT-4                           & 9841   & 5952 & 60881 & 6305  & 0.195 & 0.623       & 0.610       & 0.616      & 0.673 & 0.625 & 0.648    \\
\cmidrule(lr){1-13}
\multirow{2}{*}{Codex}         & GPT-3.5                        & 5154   & 642  & 24860 & 20838 & 0.505 & 0.889       & 0.198       & 0.324      & 0.320 & 0.161 & 0.214      \\
& GPT-4                           & 16609  & 4189 & 21313 & 9383  & 0.505 & 0.799       & 0.639       & 0.710      & 0.733 & 0.635 & 0.681      \\
\cmidrule(lr){1-13}
\multirow{2}{*}{GPT-3.5}       & GPT-3.5                        & 3582   & 211  & 11880 & 13383 & 0.584 & 0.944       & 0.211       & 0.345      & 0.303 & 0.172 & 0.220      \\
& GPT-4                           & 10724  & 1509 & 10582 & 6241  & 0.584 & 0.877       & 0.632       & 0.735      & 0.749 & 0.643 & 0.692      \\
\bottomrule
\end{tabular}
\begin{itemize}
    \item The column ``Raw Prec.'' is defined as (TP + FN) / Total, the precision of assertion generated by LLMs
\end{itemize}
\vspace{-0.4cm}
\end{table*}

\subsubsection{Trade-offs between precision and recall}
Table~\ref{tab: overallwithoutfiltering} shows the effectiveness of \technique{} without I/O example filtering for incorrect postconditions.
Compared to the results with I/O example filtering for incorrect postconditions (Table~\ref{tab: overallwithfiltering}), we observe a maximum precision increase of 0.302. 
The only difference between the two tables is that Table~\ref{tab: overallwithoutfiltering} retains many incorrect postconditions initially filtered out during I/O example filtering.
These incorrect postconditions result in the removal of both correct and incorrect assertions. 
Specifically, for the assertion generated by GPT-3.5 and the postcondtions generated by GPT-4 (the last rows in Table~\ref{tab: overallwithfiltering} and Table~\ref{tab: overallwithoutfiltering}), the reduction proportion of false positive (FP) assertions ranges from 62\% (1 - 1509 / 3950) to 79\% (1 - 211 / 4684), and the reduction proportion of true positive assertions ranges from 32\% to 79\%.
Although the reduction in both TP and FP assertions leads to an increase in precision, it also substantially increases FN assertions, thus leading to low recall values.
For instance, taking GPT-3.5 as a PostGen model, the highest recall is only 0.211, indicating the removal of the most correct assertions.

The comparison underscores that I/O example filtering for incorrect postconditions substantially reduces FN assertions (ranging from 39\% to 368\%) while increasing a small portion of FP ones. 
Using F1 scores as an evaluation metric for trade-offs, I/O example filtering for incorrect postconditions can improve the F1 scores by 70\% on average of GPT-3.5 and GPT-4.
Specifically, the improvement is 130\% and 11\% when the PostGen model is GPT-3.5 and GPT-4, respectively.



\subsubsection{Impact on the weighted metrics}
Table~\ref{tab: overallwithfiltering} also shows the results weighted across various problems in the HumanEval dataset.
We find that weighted metrics are lower than unweighted scores in most combinations of the AssertGen and PostGen models (except the combination of InCoder and GPT-4).
Such a difference arises from the distribution of assertion numbers. 
As shown in Figure~\ref{fig.histplot_unique}, these models tend to generate more assertions on problems that they can address, thus making unweighted metrics more positive.

Additionally, we noticed that the effectiveness of CodeGen decreased more than other models.
Our investigation shows that in 26 problems, CodeGen fails to generate any correct assertions, while the other three models fail in only 5, 3, and 3 problems, respectively.
Such an imbalance of CodeGen's results mainly leads to the decrease of CodeGen's ineffectiveness in weighted metrics.

%% file: evaluation/rq5.tex
\subsection{\textbf{RQ5:} How effective is \technique{} in improving the effectiveness of code generation for LLMs?}

\begin{table*}[t]

    \centering
    \footnotesize
    \caption{Overall effectiveness of \technique{} in improving the effectiveness of code generation}
    \label{tab:effectcodegen}
    \begin{threeparttable}
    \begin{tabular}{@{} c|ccc|ccc|ccc|cccc @{}} 
        \toprule
\multirow{3}{*}{Model} & \multicolumn{3}{c|}{\multirow{2}{*}{Baseline}} & \multicolumn{3}{c|}{\multirow{2}{*}{CodeT (raw)}} & \multicolumn{6}{c}{CodeT + \technique{}}                                                  \\
\cmidrule(lr){8-13}
& \multicolumn{3}{c|}{}                          & \multicolumn{3}{c|}{}                             & \multicolumn{3}{c|}{GPT-3.5 as PostGen} & \multicolumn{3}{c}{GPT-4 as PostGen} \\
        \cmidrule(lr){2-4}  \cmidrule(lr){5-7}
        \cmidrule(lr){8-10}
        \cmidrule(lr){11-13}
        & @1 & @10 & @100  & @1 & @2 & @10 & @1 & @2 & @10 & @1 & @2 & @10\\
        \midrule
        CodeGen &   29.7\% & 50.3\%  & 73.7\%    &   35.8\% & 43.0\%  & 61.6\% &   39.4\% & 47.5\%  & 62.0\% &   40.8\% & 47.5\%  & 64.4\%\\

        \midrule
        InCoder   & 16.4\% & 28.3\%  & 47.5\%&     17.7\% & 22.3\%  & 36.4\% &     23.8\% & 29.5\%  & 39.5\% &     24.5\% & 30.3\%  & 41.7\%   \\

        \midrule
        Codex  &   47.0\% & 74.9\%  & 92.1\%    &   62.9\% & 75.2\%  & 85.0\% &   64.4\% & 77.4\%  & 87.1\% &   67.0\% & 77.6\%  & 87.3\%\\

        \midrule
        GPT-3.5  &     51.2\% & 77.1\%  & 92.0\%    & 65.2\% & 72.7\%   & 85.7\%&     66.8\% & 74.9\%  & 85.7\%    & 66.9\% & 75.7\%   & 86.8\%\\
        
        \bottomrule
    
    \end{tabular}
\end{threeparttable}
\vspace{-0.3cm}
\end{table*} 

To further explore the benefits of detecting incorrect assertions, we investigate the benefits of detecting incorrect assertions in improving the effectiveness of code generation. 
Here we apply the idea of CodeT~\cite{chen2022codet}. The idea of CodeT is to generate assertions simultaneously when a LLM generates code. By designing a voting mechanism (for the same problem, multiple assertions and generated code execute each other to see the number of passes), the effectiveness of generating code from an LLM can be improved.
Similar to CodeT, we also use Pass@1, Pass@2, and Pass@10 to evaluate the effectiveness of DeCon in improving the effectiveness of code generation.

\subsubsection{Methodology}
For each LLM in generating code, we generate 1, 2, and 10 candidate programs for each problem in HumanEval. 
We first employ CodeT to directly use assertions generated by LLMs to improve the effectiveness of LLMs in generating code.
Then, in the pipeline of CodeT, we use \technique{} to remove incorrect assertions from all assertions generated by LLMs to see the improvement of LLMs' effectiveness in generating code.

\subsubsection{General Results}
As shown in Table~\ref{tab:effectcodegen}, using assertions filtered by \technique{} can improve the effectiveness of CodeT in generating code on all LLMs.
Specifically, for the postconditions generated by GPT-3.5, \technique{} improves the average Pass@K of four models by 3.2\%, 4.0\%, and 1.4\% in terms of Pass@1, Pass@2, and Pass@10, respectively. 
As for the postconditions generated by GPT-4, \technique{} improves the average Pass@K of four models by 4.4\%, 4.5\%, and 2.9\% in terms of Pass@1, Pass@2, and Pass@10, respectively. 

Comparing the Pass@K results of CodeT only and CodeT+\technique{}, we show that: (1) \technique{} can effectively improve the effectiveness of code generation for LLMs; (2) \technique{} with GPT-4 as the PostGen model is more effective than \technique{} with GPT-3.5 as the PostGen model, indicating the generalization ability when adopting further approaches for postcondition generation.



%% file: evaluation/rq6.tex
\subsection{\textbf{RQ6:} How effective is \technique{} in finding faults?}
\begin{table}[t]

    \centering
    \caption{Overall effectiveness of \technique{} on fault findings}
    \vskip 0.15in
    \vspace*{-1.5ex}
    \label{tab: faultfinding}
    \begin{threeparttable}
    \begin{tabular}{@{} cc|cccc @{}} 
        \toprule
        TestGen & PostGen
        & Total & With E.F. & Without E.F. &\\
        \midrule
        \multirow{2}{*}
        {CodeGen} & GPT-3.5 &     112 & 106  & 23  \\
        & GPT-4   &   112 & 110  & 80   \\
        
        \midrule
        \multirow{2}{*}{InCoder} &  GPT-3.5  & 144 & 139  & 36  \\
        & GPT-4   &   144 & 143  & 110  \\

        \midrule
        \multirow{2}{*}{Codex} &  GPT-3.5 &  153 & 150  & 36  \\
        & GPT-4   &   153 & 152  & 114 \\

                \midrule
        \multirow{2}{*}{GPT-3.5} &  GPT-3.5 &     152 & 146  & 31  \\
        & GPT-4   &   152 & 152  & 113   \\
  \midrule
        \multirow{2}{*}{Average} &  GPT-3.5 &     140.25 & 135.25  & 31.5  \\
        & GPT-4   &   140.25 & 139.25  & 104.25   \\

        \bottomrule
    
    \end{tabular}
\end{threeparttable}

\begin{itemize}
    \item E.F. indicates ``Example Filtering'' (i.e., postconditions that are filtered by I/O examples)
\end{itemize}
\end{table} 
To further show the effectiveness of \technique{}, we conduct fault-finding experiments to show the fault-finding ability after detecting and removing incorrect assertions.
The fault-finding experiment shows that although DeCon might filter out correct assertions, the fault-finding ability of the rest of the correct assertions decreases slightly. 
Compared with the fault-finding ability of assertions filtered by postconditions (without filtering by I/O examples), the fault-finding ability of assertions filtered by postconditions (with filtering by I/O examples) is improved.

\subsubsection{Methodology}
In this section, we compare the effectiveness of assertions generated by LLMs without filtering by \technique{} and those filtered by \technique{} from the perspective of fault-finding.
For all four LLMs, we combine the faults that are found by the correct assertions (based on our ground truth given in the HumanEvalFix dataset) as the ground-truth faults.
Then, we evaluate how many problems (out of the 164 ones in total) in the HumanEvalFix dataset whose ground-truth faults can be found by the assertions with/without \technique{}'s example filtering (i.e., filtering incorrect postconditions based on the I/O examples).


\subsubsection{General Results}
Table~\ref{tab: faultfinding} shows the overall effectiveness of \technique{} in fault finding.
The Column ``Total'' refers to the total number of faults that can be found in the correct assertions generated by each LLM. For some problems, the assertions generated by the LLM cannot find faults.
We find that the correct assertions generated by CodeGen have the worst fault-finding ability, finding faults from 114 out of the 164 problems, while the other three models have better fault-finding abilities, among which Codex and GPT-3.5 can find 153 and 152 out of the 164 problems, respectively.
The Column ``With E.F.'' refers to the effectiveness of assertions filtered by postconditions; these postconditions are filtered by I/O examples.
On average, over various AssertGen models, 96.4\% (135.25/140.25) and 99.3\% (139.25/140.25) of the problems with ground-truth faults can be found by assertions generated by \technique{} with I/O example filtering on postconditions generated by GPT-3.5 and GPT-4, respectively.

On the contrary, postconditions without I/O example filtering substantially decrease \technique{}'s effectiveness in fault-finding.
The Column ``Without E.F.'' refers to the effectiveness of assertions filtered by postconditions; these postconditions are not filtered by I/O examples.
Specifically, there are only 35\% and 82\% of the correct assertions that are retained after being filtered by postconditions (generated by GPT-3.5 and GPT-4, respectively) without I/O examples filtering.
Thus, the remaining correct assertions can only detect 22.5\% and over 74.3\% of the problems with faults, respectively.
The decreasing effectiveness is due to the incorrect postconditions that can be filtered by I/O examples.
These incorrect postconditions remove a large portion of TP assertions, which contribute to the effectiveness of fault-finding (finding more than 96\% of problems with faults).

The preceding evaluation results indicate that: (1) assertions generated by \technique{} can find most faults when compared to all correct assertions that are not filtered by postconditions; 
(2) Compared to not using I/O examples to filter out the incorrect postconditions generated by LLMs, \technique{} can find 73.9\% (96.4\%-22.5\%) and 25\% (99.3\%-74.3\%) more fault findings with the postconditions generated by GPT-3.5 and GPT-4, respectively.

\subsubsection{Compare with I/O examples in docstring}
We have conducted the baseline experiment that can find faults by using only I/O examples extracted from the function's docstring description. 
The number of problems found to be faulted is 152, consistent with DeCon. While our experimental results confirm the consistency in the number of faults with DeCon, they also underscore the complementarity of utilizing both I/O examples and DeCon. Notably, among 12 problems that cannot be found with faults using I/O examples in the docstring, DeCon managed to identify 10 out of the 12 problems.



%% file: discussion.tex
\section{discussion}\label{sec:discuss}


\subsection{Selection Criteria of PostGen Models}
The reason why we did not use CodeGen, InCoder, and Codex to generate postconditions is that we tried a lot of prompts to make these models generate formal and executable (i.e., assert ``xxx'') postconditions, but none of these models were able to generate them successfully.
The existing work~\cite{yao2023leveraging, endres2023formalizing} uses GPT-3.5 and GPT-4 to generate prompts for postconditions. We have successfully used GPT-3.5 and GPT-4 to generate formal and executable postconditions for problems in HumanEval. Therefore, we selected GPT-3.5 and GPT-4 as the models to generate postcondition.

\subsection{Selection Criteria of TestGen Models}
The work of CodeT~\cite{chen2022codet} uses CodeGen, InCoder, and Codex to generate test cases. We use the same prompt as this work to generate assertions for CodeGen, InCoder, and Codex. To compare whether generating postconditions for the same model can improve the performance of the test cases generated by the model, in addition to the three models mentioned above, we use the same prompt to generate test cases for GPT-3.5. We did not use GPT-4 to generate test cases here, mainly because GPT-4 is too expensive.

\subsection{How effective of existing LLMs in generating assertions with I/O examples?}
To provide a fair comparison between the assertion and postcondition generation~\footnote{We use I/O examples to filter out incorrect postconditions, and we do not use I/O examples in assertion generation }, we conduct a comparative experiment for GPT-3.5 (i.e., giving i/o examples when generating assertions). We found that there was almost no decrease in the proportion of incorrect assertions generated by LLMs or the proportion of false positives that our approach could reduce. Experimental results show that static use of I/O examples is not enough and dynamic filtering is required.

\subsection{Compared with DL-based Assertion Generation Approaches}
We do not use assertion generation approaches based on deep learning to generate assertions for two reasons. First, current assertion generation approaches based on deep learning are based on a limited training set (training data and program language are all limited) and cannot be generalized to the HumanEval dataset. Second, a study~\cite{Tufano2022assert} has shown that the effectiveness of LLM-based assertion generation is higher than that of deep learning-based approaches.

Traditional deep-learning approaches require specific training sets. We have tried ATLAS and TOGA, but they cannot generate correct assertions for the functions in HumanEval.
The traditional unit test case generation approaches generate the assertion by executing the method to be tested, so it can never find bugs, and it isn't significant to reduce its false positives.






%% file: threats.tex
\section{Threats to Validity}\label{sec:threats}
The first threat is that DeCon might not be generalized to the functions under test with the input parameters or return value of non-basic types.
Currently, we have only validated the effectiveness of DeCon on HumanEval. We plan to extend DeCon to real-world development scenarios in our future work.
The second threat is that the improvements over the baselines for RQ5 and RQ6 to be statistically not too significant, so it is difficult to assess the impact of DeCon for code generation and fault finding.
The third threat is that the results on the baseline and the improvements over CodeT(raw) for RQ5 could potentially indicate a data leak.


%% file: relatedwork.tex
\section{related work}\label{sec:relatedwork}

\subsection{Postcondition Generation}

Yao~\cite{yao2023leveraging} et al. present using GPT4 to generate postconditions and using static analysis to synthesize invariants, assertions, and other proof structures for a Rust-based formal verification framework.
They find that LLMs demonstrate impressive logical ability in generating postconditions. 
Endres et al.~\cite{endres2023formalizing} present LLM4nl2post to transform informal natural language to formal method postconditions, expressed as program assertions. Endres et al. introduce and validate metrics to measure and compare different LLM4nl2post approaches, using the correctness and discriminative power of generated postconditions. Then, Endres et al. perform qualitative and quantitative methods to assess the quality of LLM4nl2post postconditions, finding that they are generally correct and able to discriminate incorrect code. Endres et al. find that LLM4nl2post via LLMs has the potential to be helpful in practice; specifications generated from natural language were able to catch 70 real-world historical bugs from Defects4J.

\subsection{Assertion Generation}
\subsubsection{Traditional assertion generation approaches} Traditional assertion generation approaches (that help detect only crashing faults or regression faults, and are incapable of detecting non-crashing faults in the current version in the absence of a previous version) can automatically generate assertions with two main categories.   
(1) Capture and assert~\cite{xie06ecoop}. For example, Randoop~\cite{Randoop24} and EvoSuite~\cite{Evosuite} create assertions based on capturing and asserting the return values of all non-void-return methods of the method sequence in the generated test input; EvoSuite further reduces these assertions based on mutation testing~\cite{mutation1,mutation2}.
(2) Differential testing~\cite{Evans2007assertion}. For example, DiffGen~\cite{Taneja2008assertion} generates assertions from runs on two different versions of a class by checking the equality/equivalence of method-call return values and receiver object states from the two versions.

\subsubsection{DL-based assertion generation approaches} In recent years, many approaches~\cite{dinalla2022toga,yu2022automated,deepAssert} take a test method without any assertion (i.e., test input only) along with its focal method (i.e., the method under test).  
ATLAS~\cite{deepAssert} is the first approach to use deep learning to generate assertions through testing methods and their focus methods.
Yu et al.~\cite{yu2022automated} first tried to use information retrieval (IR) in assertion generation, and proposed an IR-based method, including IR-based assertion retrieval technology and retrieved assertion adaptation technology. In addition, they proposed an integrated approach that combines IR-based approaches with DL-based approaches (e.g., ATLAS) to further improve efficiency.
Tufano et al.~\cite{Tufano2022assert} proposed a method to generate precise assertion statements based on the sequence-to-sequence converter model. This method can predict correct assertions in 62\% of the cases in the first attempt. 
TOGA~\cite{dinalla2022toga} is a unified neural method based on a converter, which is used for context inference exception and assertion test prediction based on the focus method.

%% file: conclusion.tex
\section{conclusion}\label{sec:conclusion}

In this paper, we have conducted an empirical study on the quality of assertions generated by four LLMs on the HumanEval and found that 62.4\% of generated assertions are incorrect.
Based on only a few I/O examples besides the docstring and the target function signature for the target problem, we have proposed an approach named \technique{} to detect whether a generated assertion is incorrect via postcondition generation.
Experimental results have shown that \technique{} can detect more than 64\% incorrect assertions generated by LLMs, and \technique{} can also improve the effectiveness of LLMs in the tasks of code generation and fault-finding.

\section{Data Availability}
We open-source our source code, dataset, and evaluation results on our anonymous website~\cite{sourcedata}.